\documentclass[notitlepage,twocolumn,nofootinbib]{revtex4-1} 

\usepackage{amsmath}         
\usepackage{amssymb}         
\usepackage{amsfonts}        
\usepackage{graphicx}        

\usepackage{lipsum}         
\usepackage{url}        

\renewcommand{\vec}[1]{\mathbf{#1}} 

\graphicspath{{figures/}} 

\begin{document}

\title{On-Shell Mediators and Top--Charm Dark Matter Models for the Fermi-LAT Galactic Center Excess}

\author{Arvind Rajaraman}
\email[]{\texttt{arajaram@uci.edu}}
\affiliation{Dept.\ of Physics \& Astronomy, University of California, Irvine, CA 92697}

\author{Jordan Smolinsky}
\email[]{\texttt{jsmolins@uci.edu}}
\affiliation{Dept.\ of Physics \& Astronomy, University of California, Irvine, CA 92697}

\author{Philip Tanedo}
\email[]{\texttt{flip.tanedo@uci.edu}}
\affiliation{Dept.\ of Physics \& Astronomy, University of California, Irvine, CA 92697}

\date{\today}

\begin{abstract}
An excess in $\gamma$-rays from the galactic center observed by the \textsc{fermi} telescope has been proposed as a possible signal of dark matter annihilation. 
Recently, the \textsc{fermi} collaboration showed that systematic errors broaden the range of spectral shapes for this excess.
We demonstrate fits to this range for (1) flavor-violating annihilations to top--charm pairs and (2) annihilations to on-shell bosonic mediators which decay to Standard Model quarks in a boosted frame.
Annihilation of 40 -- 100 GeV DM to pairs of spin-1 mediators provide a good fit to the \textsc{fermi} spectrum with a normalization consistent with a thermal relic.  Top--charm modes and annihilation to three pseudoscalar mediators can fit the spectral shape but typically require non-thermal annihilation cross sections.

\end{abstract}

\maketitle

\section{Introduction}

Analyses of the data from the \textsc{fermi} Space Telescope show an excess of $\gamma$-rays with energy 1--10 GeV emanating from the center of the galaxy~\cite{Goodenough:2009gk, Hooper:2010mq, Boyarsky:2010dr, Abazajian:2012pn, Hooper:2011ti,Linden:2012iv,Abazajian:2012pn, Gordon:2013vta,Abazajian:2014fta,Daylan:2014rsa,Zhou:2014lva,Calore:2014xka} and at high galactic latitudes~\cite{Hooper:2013rwa, Okada:2013bna, Huang:2013pda}. More recent studies, including one by the \textsc{fermi} collaboration~\cite{Murgia:2014}, include estimates of the systematic uncertainties on this excess~\cite{Zhou:2014lva, Calore:2014xka, Calore:2014nla} and have broadened the range of allowed spectra.
Proposed astrophysical explanations exist~\cite{Abazajian:2010zy, Petrovic:2014xra,Macias:2013vya,Petrovic:2014uda,Carlson:2014cwa}, but it is not yet clear if any single astrophysical mechanism can account for the excess on all angular scales.

An intriguing alternate origin for the excess is the annihilation of dark matter (DM) into Standard Model (SM) final states which later shower to photons.
Early DM fits to the excess preferred $\mathcal O(40)$ GeV DM annihilating into $b$ pairs or $\mathcal O(10)$ GeV DM annihilating into lepton pairs with a cross section compatible with that required for a thermal relic; see~\cite{Izaguirre:2014vva,Berlin:2014tja,Alves:2014yha,Lacroix:2014eea} for phenomenological models. Including the systematic uncertainties from more recent studies~\cite{Calore:2014nla, Murgia:2014} allows a range of $\mathcal O(10-100)$ GeV DM masses and several options for SM final state pairs~\cite{Agrawal:2014oha} (a similar range of masses was noted in~\cite{Abazajian:2012pn}) and can alleviate bounds from dwarf spheroidals~\cite{Ackermann:2015zua}.

When DM annihilates into two SM particles, the $\gamma$-ray spectrum is purely determined by the identity of the SM final state and the dark matter mass. 
If these interactions are mediated by heavy particles, the couplings required to reproduce the $\gamma$-ray excess are constrained by monojet~\cite{Alves:2014yha} and direct detection experiments (see e.g.~\cite{DelNobile:2013sia}). 
This suggests a dark sector that contains light mediators. In the limit where these mediators are lighter than the dark matter, DM can annihilate to on-shell mediators, and the annihilation rate is independent of the mediator coupling to the SM~\cite{ArkaniHamed:2008qn}. 
In this way, one may parametrically avoid the simplest direct detection and collider bounds while preserving the \textsc{fermi} signal. This scenario was explored recently in~\cite{Boehm:2014bia, Ko:2014gha, Abdullah:2014lla, Martin:2014sxa} for annihilation into quarks, \cite{Kaplinghat:2015gha} for annihilation into electrons, and \cite{Elor:2015tva} for the case of multistep cascades introduced in~\cite{Mardon:2009rc}.

\begin{figure} 
\includegraphics[width=.47\textwidth]{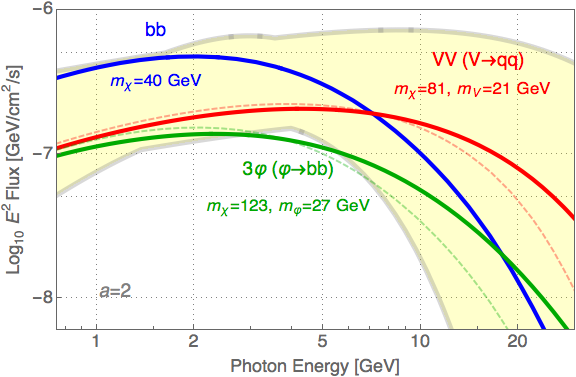}
\caption{Benchmark $\gamma$ spectra for $s$-wave Dirac DM annihilation at the thermal relic cross section: $b\bar b$ (blue), on-shell vectors going to light quarks (red), on-shell pseudoscalars going to $b\bar b$ (green). The yellow band is the envelope in (\ref{eq:env}). Dashed lines correspond to the expected spectral shape without the boost from on-shell mediators (see text).
\label{fig:sampleplots}}
\end{figure}

In this note, we compare the excess presented by the \textsc{fermi} collaboration~\cite{Murgia:2014} to three new types of dark matter models. In the first, dark matter annihilates to a $t\bar{c}$ final state. In the second, dark matter annihilates to two vector mediators which each decay to two quarks. In the third, dark matter annihilates to three pseudoscalars, which each decay to two quarks.
For each class of models, we use the \textsc{pppc} package~\cite{Cirelli:2010xx, Ciafaloni:2010ti} for \textit{Mathematica}~\cite{Mathematica10} to generate spectra. We  compare these to the spectra presented by the \textsc{fermi} collaboration, and find the allowed parameter space of the model. 

We find that many of these models are viable; for instance, as shown in Fig.~\ref{fig:sampleplots}, the spectra are well fit by the decay to vector mediators with a somewhat larger DM mass and where the dark matter has the correct thermal annihilation cross section. We also show that the shapes of the spectra from the two other classes are compatible with the spectra presented by the \textsc{fermi} collaboration, but require a non-thermal component to the relic density.
Figs.~\ref{fig:spec:tc:v:bb},~\ref{fig:spec:vv},~and~\ref{fig:spec:3p} summarize our results. We show that
\begin{enumerate}
	\item Top--charm annihilation modes provide viable spectra that are very similar to $b\bar b$.
	\item Vector mediators with universal quark couplings are consistent with the spectrum and thermal relic abundance for DM masses between 40 -- 100 GeV.
	\item Pseudoscalar mediators have a range of masses which fit the spectral shape but typically require $\langle \sigma v\rangle$ larger than the thermal relic value.
\end{enumerate}
These qualitative results are only weakly dependent on the mediator mass.

\section{$\gamma$ Spectra from DM Annihilation}
\label{sec:spectra}

The spectrum of the photon flux at Earth $d\Phi/dE$ from the annihilation of DM $\chi$ of mass $m_\chi$ and with galactic density profile $\rho_\text{\textsc{dm}}(r)$ is
\begin{align}
	\frac{d\Phi}{dE}
	&= 
	\int 
	\frac{r^2 dr d\Omega}{4\pi r^2} \frac{\rho_\text{\textsc{dm}}^2}{m_\chi^2} \frac{\langle \sigma v \rangle}{\eta} \frac{dN}{dE}
	\equiv
	\frac{\langle \sigma v \rangle}{4\pi\eta m_\chi^2} \frac{dN}{dE}
		J, 
		\label{eq:photon:flux}
\end{align}
where $\langle \sigma v\rangle$ is the thermally averaged cross section for DM annihilation and $dN/dE$ is the spectrum of photons from the final states and is encoded in \textsc{pppc}~\cite{Cirelli:2010xx}. 
The integral of the squared DM density profile is $J=1.08\times 10^{23}\text{ GeV}^2/\text{cm}^5$ assuming an NFW profile~\cite{Navarro:1995iw, Navarro:1996gj} with $\gamma=1.2$ following the $15^\circ\times 15^\circ$ region of interest about the galactic center and fit by the \textsc{fermi} analysis~\cite{Murgia:2014}.
$\eta = 2\,(4)$ for Majorana (Dirac) dark matter\footnote{%
These values have different origins.
For Majorana DM, this corrects a double counting in the thermally averaged initial state phase space in $\langle \sigma v \rangle$. 
For Dirac DM, this accounts for annihilation being proportional to $n_\chi n_{\bar\chi} = \rho_\text{\textsc{dm}}^2/4m_\chi^2$.
	See, e.g.~\cite{Gondolo:1990dk}.
}.
One typically assumes that annihilation is $s$-wave and consistent with the rate required for a thermal relic~\cite{Steigman:2012nb, Feng:2008mu},
\begin{align}
\langle \sigma v\rangle \approx 1.1\,\eta \times 10^{-26}\text{ cm}^3/\text{s}.
\label{eq:thermal:relic}	
\end{align}
For the remainder of this paper we assume Dirac DM so that $\eta = 4$.
In the case of on-shell mediators the spectra $dN/dE$ are (1) smeared out due to the boosting of the SM states produced from mediator decay (2) enhanced because each annihilation produces a multiplicity of final states according to the number of mediators; see~\cite{Abdullah:2014lla} for details.

We compare the DM annihilation spectra of on-shell mediator models to those observed by the \textsc{fermi} telescope. We refer to~\cite{Agrawal:2014oha} for a $\chi^2$ fit to models with DM annihilation to pairs of SM final states. However, the analyses in~\cite{Calore:2014nla,Calore:2014xka} notwithstanding, it is difficult to quantify systematic errors in this excess. We thus take a complementary approach and compare our spectra to the power law with exponential cutoff fits from \textsc{fermi} collaboration analysis presented in~\cite{Murgia:2014}. 
	We use the four spectra $E^2 d\Phi_i/dE$ in that analysis to define an envelope that is meant to estimate the systematic error on the excess:
\begin{equation}
		\begin{aligned}
		\frac{d\Phi_{\text{min}}(E)}{dE} &= a^{-1}\text{min}_i\left\{ \frac{d\Phi_i(E)}{dE}\right\}
		\\
		\frac{d\Phi_{\text{max}}(E)}{dE} &= a\;\text{max}_i \left\{ \frac{d\Phi_i(E)}{dE}\right\},
	\end{aligned}
	\label{eq:env}
\end{equation}
	where $a$ a scaling of the envelope size that we choose to be 2 in this paper\footnote{This is comparable to the range $\mathcal J\in [0.14,4.0]$ in \S3.2 of~\cite{Agrawal:2014oha}.}; 
	see Fig.~\ref{fig:sampleplots}. 
We quantify the fit of DM annihilation spectra relative to the envelope (\ref{eq:env}) in two ways:
\begin{enumerate}
	\item For a fixed envelope normalization $a$, we specify the range of annihilation cross sections $\langle \sigma v\rangle$ that fall within the envelope. This makes it simple to compare to the target thermal relic cross section. 
	\item For a given final state spectrum $dN/dE$, we specify the minimum envelope normalization $a$ that may accommodate this shape. This quantifies the quality of the spectral shape fit. We present these plots in Appendix~\ref{app:shape:fits}.
\end{enumerate}
We emphasize that this is an estimate of the parameter space for the on-shell mediator scenario that is complementary to the conventional $\chi^2$ analysis in~\cite{Agrawal:2014oha}.

\section{Annihilation to top-charm}
\begin{figure}
\includegraphics[width=.45\textwidth]{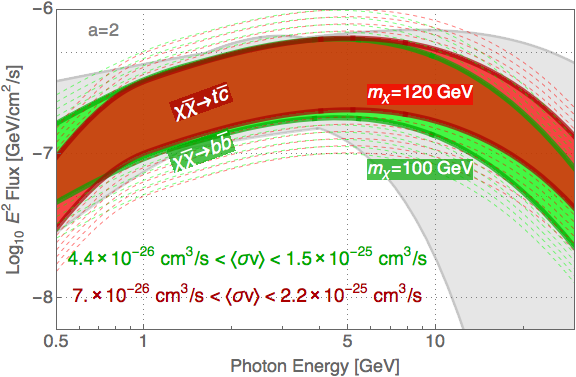}
\caption{Comparison of the $t\bar c$ (red) and  $b\bar b$ (green) spectra for a choice of parameters.
	The bands show normalizations that fit the envelope~(\ref{eq:env}), corresponding cross sections are given: the upper (lower) range corresponds to the lighter (heavier) DM, shown in green (red). 
	Dashed lines show the shape of nearby normalizations.
\label{fig:spec:tc:v:bb}}
\end{figure}

\begin{figure}
\includegraphics[width=.4\textwidth]{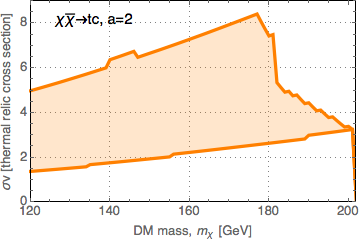}
\caption{Annihilation cross section required for $t\bar c$ mode to fit within the envelope~(\ref{eq:env}). 
\label{fig:tc:x}
}
\end{figure}

Previous analyses of DM annihilation to pairs of SM particles  focus on flavor-conserving fermion pairs or boson~\cite{Agrawal:2014una}. One may also consider flavor-violating modes. For final state particles  $i$ and $j$ with mass splitting, $\Delta m_{ij}^2=m_i^2 -m_j^2$, the injection energies differ,
\begin{align}
	E_i &= \frac{4m_\chi^2 + \Delta m_{ij}^2}{4m_\chi}.
\end{align}
The resulting photon spectrum is a combination of two spectra with different characteristic energy scales. The mass differences for first and second generation quarks are too small for this to be an appreciable effect. Further, transitions between $b$-quarks and lighter quarks are tightly constrained from the $B$-factories~\cite{Bevan:2014iga, Agrawal:2014aoa}. Thus the most plausible charge-neutral combination is a $t\bar c$ interaction~\cite{Kamenik:2011nb, Calibbi:2015sfa}; this has a range of UV motivations and collider signatures that are unique from the flavor-conserving case~\cite{Blanke:2013uia, Craig:2012vj, Chen:2013qta}. We leave lepton flavor-violating modes for future work~\cite{Galon:2015}.

Note that the $t\bar t$ spectrum is generally a difficult fit to the \textsc{fermi} excess due to the hardness of the spectrum~\cite{Agrawal:2014oha}; $t\bar c$ alleviates this since  it only requires $2m\chi \gtrsim 180$ GeV rather than $2m_t$ and this can better fit the envelope~(\ref{eq:env}). We show this by comparing to a $\chi\bar\chi \to b\bar b$ spectrum in Fig.~\ref{fig:spec:tc:v:bb}. Note the similarity of the two spectra for the choice of parameters.
Fig.~\ref{fig:tc:x} shows the annihilation cross section required to fit in the envelope with $a=2$; one is restricted to values of $m_\chi$ near its lowest allowed value to be near the thermal cross section.

\section{Annihilation to Spin-1 Mediators}

\begin{figure}
\includegraphics[width=.38\textwidth]{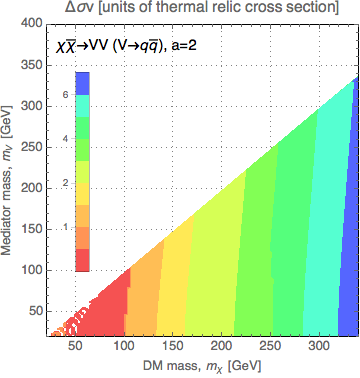}
\caption{Separation of the $\langle \sigma v\rangle$ required to fit in the \textsc{fermi} envelope from the thermal value in units of the thermal value for the case of annihilation to two spin-1 mediators which decay to light quarks. The range $m_\chi \in [50,100]$ GeV is consistent with a thermal relic. \label{fig:vqq:x}}
\end{figure}

The leading $s$-wave contribution to DM annihilation to on-shell spin-1 mediators is $\chi\bar\chi\to VV$, where $V$ may be either CP even or odd~\cite{Boehm:2014bia, Abdullah:2014lla, Martin:2014sxa}. With the modest assumption of minimal flavor violation, one expects such a mediator to couple universally to all quark flavors. Here we present results for universal couplings to light quarks; the modification from the $b$ quark contribution is a percent level correction. For completeness we present results a $b$-philic vector in Appendix~\ref{app:MFV:violating}. 

\begin{figure}
\includegraphics[width=.45\textwidth]{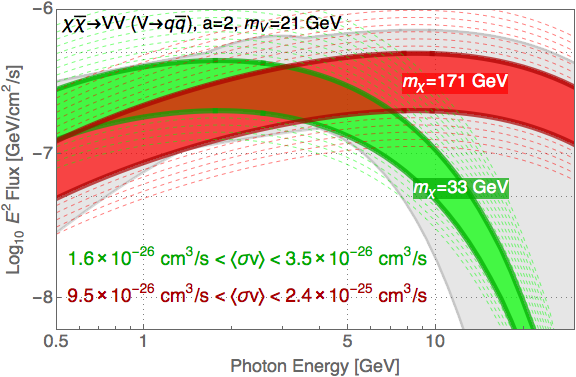}
\caption{Extreme spectra for annihilation to on-shell, universally coupled vector mediators.   Conventions as in Fig.~\ref{fig:spec:tc:v:bb}. \label{fig:spec:vv}}
\end{figure}

As noted above, the on-shell mediator smears out the photon spectrum. In the non-relativistic limit, kinematics restrict the $V$s to be mono-energetic. However, the photon spectra from $V\to q\bar q$ must be convoluted with a box spectrum that encodes the boost from the mediator rest frame~\cite{Abdullah:2014lla}. This smears out the spectrum, as demonstrated by the dashed lines in Fig.~\ref{fig:sampleplots}. Fig.~\ref{fig:spec:vv} plots the shapes of two extreme examples of allowed DM masses in this scenario. The range of normalizations that allow each shape to fit the envelope~(\ref{eq:env}) are listed for comparison with (\ref{eq:thermal:relic}).

This scenario is able to accommodate a thermal relic within the \textsc{fermi} envelope~(\ref{eq:env}). We note that this class of mediators is constrained theoretically by anomaly cancellation~\cite{Dobrescu:2014fca}. See~\cite{Hooper:2014fda} for a recent exploration of viable $Z'$ models.

\section{Three pseudoscalar mediators}

For spin-0 mediators, the leading $s$-wave on-shell annihilation mode comes from the annihilation to three scalars~\cite{Abdullah:2014lla}. We consider the simplest case, $\chi\bar\chi \to 3 \varphi$, for a pseudoscalar $\varphi$. The energy spectrum of these mediators is no longer a simple box-spectrum and must be convoluted with the photon spectra. We present expressions for calculating this mediator spectrum in Appendix~\ref{app:pseudoscalar:spectrum}. In contrast to spin-1 mediators, minimal flavor violation imposes that spin-0 mediators couple proportionally to the Yukawas so we make the approximation of an exclusive $\varphi \to b\bar b$ decay.

\begin{figure}
\includegraphics[width=.38\textwidth]{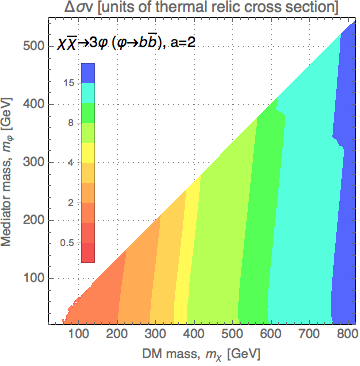}
\caption{
Separation of the $\langle \sigma v\rangle$ required to fit in the \textsc{fermi} envelope from the thermal value in units of the thermal value for the case of annihilation to three pseudoscalar mediators which decay to $b$s. The thermal cross section is generally not large enough not explain the \textsc{fermi} excess.
\label{fig:pbb:x}}
\end{figure}

\begin{figure}
\includegraphics[width=.45\textwidth]{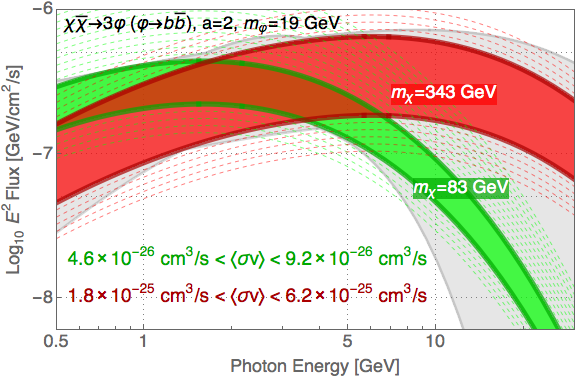}
\caption{Extreme spectra for annihilation to on-shell, $b$-philic pseudoscalar mediators.   Conventions as in Fig.~\ref{fig:spec:tc:v:bb}. \label{fig:spec:3p}}
\end{figure}

\begin{figure}
\includegraphics[width=.4 \textwidth]{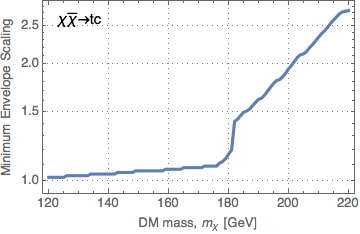}
\caption{Minimum $a$ required in (\ref{eq:env}) for $\chi\bar\chi \to t\bar c$.  The spectral shape is in tension with the envelope for $m_\chi \gtrsim 180$ GeV. \label{fig:tc:a}}
\end{figure}

We show fits for this scenario in Figs.~\ref{fig:pbb:x} and \ref{fig:spec:3p}. There is some tension with the thermal relic cross section. This can be seen in the expression for the photon flux~(\ref{eq:photon:flux}): while $dN/dE$ is three times larger than that of $\chi\bar\chi \to b\bar b$ due to the multiplicity of $b$s, the flux also decreases by $9$ due to the more sparse DM number density associated with a heavier $\chi$. This is because $m_\chi$ must be approximately three times heavier so that each $b$ has roughly the same injection energy as in the $\chi\bar\chi \to b\bar b$ case. Hence $\langle \sigma v \rangle$ must compensate for this overall $\mathcal O(1/3)$ factor. In fact, as explained in~\cite{Abazajian:2010zy}, this is still an over-estimate for comparison to the thermal relic cross section since at freeze out, the $p$-wave annihilation mode to two $\varphi$s is not significantly suppressed relative to the $3\phi$ mode. Thus, this case may require a non-thermal DM abundance, see e.g.~\cite{Baer:2014eja,Zurek:2013wia,Petraki:2013wwa}.

A template simplified model for pseudoscalar mediators and the \textsc{fermi} excess was presented in~\cite{Boehm:2014hva} with the on-shell limit considered in~\cite{Abdullah:2014lla}. Since spin-0 particles couple left- and right-chiral fermions, UV completions of this framework typically require interactions with the Higgs~\cite{Ipek:2014gua}; see recent studies in two-Higgs doublet~\cite{Berlin:2015wwa} and (N)MSSM~\cite{Cheung:2014lqa, Cahill-Rowley:2014ora, Gherghetta:2015ysa} frameworks.

\begin{acknowledgments}
\noindent This work is supported in part by the \textsc{nsf} grant \textsc{phy}-1316792. \textsc{p.t.}~is supported in part by a \textsc{uci} Chancellor's \textsc{advance} fellowship.
We are grateful to 
Kev Abazajian,
Prateek Agrawal,
Marco Cirelli,
Bogdan Dobrescu,
Marat Freytsis,
Shunsaku Horiuchi,
Manoj Kaplinghat,
G\u{o}rd\r{a}n Krnj\"ai\c{c},
Simona Murgia,
Stefano Profumo,
Tim~M.P.~Tait,
and
Hai-Bo Yu
for many enlightening discussions. 
\textsc{p.t.} thanks the Aspen Center for Physics (\textsc{nsf} grant \#1066293) where part of this work was completed. \textit{Mathematica} notebooks for this analysis are available at \url{http://git.io/hzHD}. 
\end{acknowledgments}

\appendix

\section{Shape Fits}
\label{app:shape:fits}

\begin{figure*}
\begin{center}
\includegraphics[width=.36 \textwidth]{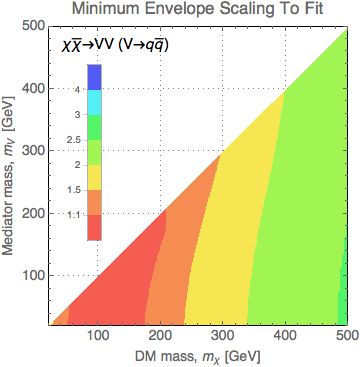}
\qquad
\qquad
\qquad
\qquad
\includegraphics[width=.36 \textwidth]{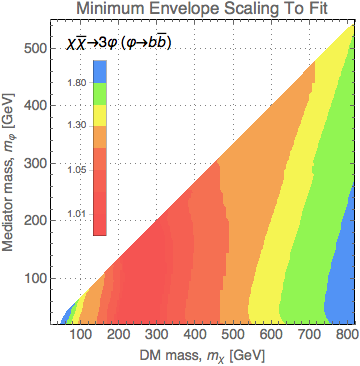}	
\end{center}
\caption{Contour plots for the minimum envelope scaling to accommodate the spectral shape of on-shell spin-1 (left) and pseudoscalar (right) mediators. Contours correspond to $a$ in (\ref{eq:env}). \label{fig:shape:fits}}
\end{figure*}

As explained in Sec.~\ref{sec:spectra}, there are two ways to quantify the fit of a particular photon spectrum. In the main text, Figs.~\ref{fig:vqq:x} and~\ref{fig:pbb:x} showed contours of $\langle \sigma v\rangle$ required for a given spectrum to fit in the envelope (\ref{eq:env}). This quantifies overall normalization. Alternately, one may quantify the fit of the spectral shape by the minimum envelope scaling $a$ for which there exists any $\langle \sigma v\rangle$ that allows the spectrum to fit the envelope. This data is plotted for the cases in the main text in Figs.~\ref{fig:tc:a} and~\ref{fig:shape:fits}.

\section{$s$-wave Pseudoscalar Spectrum}
\label{app:pseudoscalar:spectrum}

We summarize the spectrum $dN_\varphi/dE_\varphi$ of non-relativistic DM annihilation into three pseudoscalars. For simplicity we set the coupling to one. The squared amplitude is
\begin{align}
	\frac 14 \sum_{\text{spins}}\left| \mathcal M \right|^2
	&= m_\chi^2 \sum_{ABCD} \left(R_{AB} R_{CD}^*\right)_{A\neq B, C\neq D},
\end{align}
where $A,B,C,D \in \{1,2,3\}$ label the final state $\varphi$ and
\begin{align}
		R_{AB} &=
	\frac{\left(P^0_A + m_\chi\right) \left(-Q^0_A + m_\chi\right) + \vec{P}_A\cdot\vec{Q}_B}{\left(m_\varphi^2 - 2mE_A\right)\left(m_\varphi^2 - 2mE_B\right)},
\end{align}
written with respect to the 4-vectors formed from the outgoing $\varphi$ momenta $p_A$ and the incoming DM (anti-DM) momenta $k_{1(2)}$,
\begin{align}
	P_A = p_A - k_2 
	\qquad\qquad
	Q_A = k_1 - p_A.
\end{align}
The differential cross section is 
\begin{align}
	d\langle\sigma v \rangle &=  \frac{1}{4s}\left|\mathcal M \right|^2 \, \frac{1}{(2\pi)^3}\frac{p_1 dp_1}{2E_1}\frac{p_2 dp_2}{2E_2}.
\end{align}
In order to find $d\langle\sigma v\rangle/d E_\varphi = 3\, d\langle\sigma v\rangle/d E_1$, we must integrate over $dp_2$ over its allowed kinematic range. Define an energy function
\begin{align}
	\epsilon(p_1,p_2,\cos \theta) = E_1 + E_2 + E_3(p_1,p_2,\cos\theta),
\end{align}
where $\theta$ is the angle between $\vec{p}_1$ and $\vec{p}_2$, $p_{1,2}$ are lengths of 3-momenta, $E_{1,2}^2 = p_{1,2}^2 + m_\varphi^2$, and 
\begin{align}
	E_3^2 = p_1^2 + p_2^2 + 2 p_1p_2 \cos\theta + m^2.
\end{align}
The region of $dp_2$ integration is given by the condition that there exists some $|\cos \theta| \leq 1$ such that $\epsilon(p_1,p_2,\cos\theta) = \sqrt{s}=2m_\chi$. This can be visualized by plotting $\epsilon(p_1,p_2,-1)$ as a function of $p_2$ and noting that for general values of $\cos\theta$, $\epsilon$ lies above this line. There are two regions. First, when $\epsilon(p_1,0,-1) >\sqrt{s}$, then $p_2^\text{min}$ and $p_2^\text{max}$ are given by the solution to
\begin{align}
	\epsilon(p_1,p_2^\text{min,max},-1) = \sqrt{s}.
\end{align}
The second region has $\epsilon(p_1,0,\cos\theta) < \sqrt{s}$, in this case $p_2^\text{min}$ and $p_2^\text{max}$ are given respectively by the solutions to
\begin{align}
	\epsilon(p_1,p_2^\text{min},+1) &= \sqrt{s}\\
	\epsilon(p_1,p_2^\text{max},-1) &= \sqrt{s}.
\end{align}
Upon performing the $dp_2$ integration, the $\varphi$ spectrum can be found using the density of states~\cite{Fortin:2009rq},
\begin{align}
	\frac{1}{N_\varphi} \frac{dN_\varphi}{dE_\varphi} &= \frac{1}{\langle \sigma v \rangle}\frac{d\langle \sigma v\rangle}{dE_\varphi}.
\end{align}

\section{MFV-Violating Modes}
\label{app:MFV:violating}

\begin{figure*} 
\begin{center}
\includegraphics[width=.38 \textwidth]{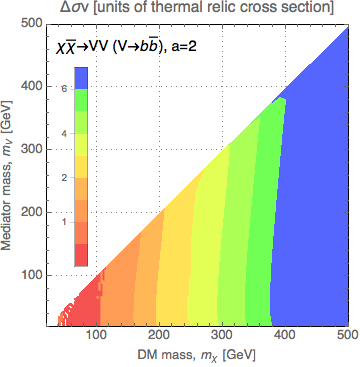}
\qquad\qquad\qquad
\includegraphics[width=.38 \textwidth]{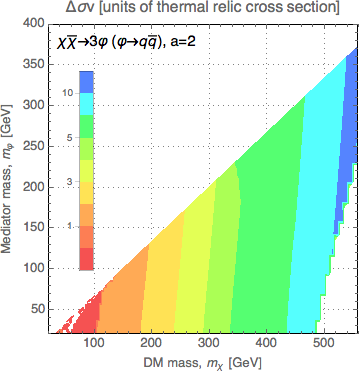}
\\
\includegraphics[width=.38 \textwidth]{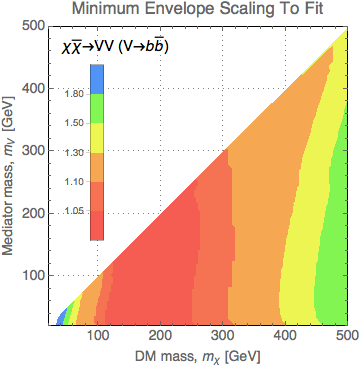}
\qquad\qquad\qquad
\includegraphics[width=.38 \textwidth]{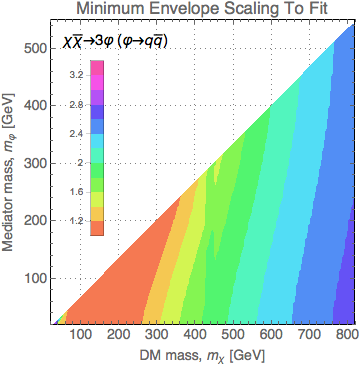}	
\\
\includegraphics[width=.4 \textwidth]{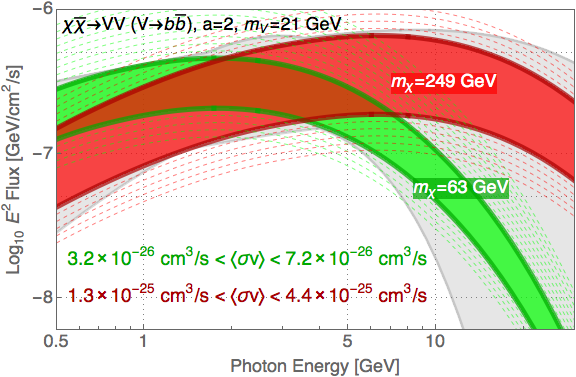}
\qquad\qquad\qquad
\includegraphics[width=.4 \textwidth]{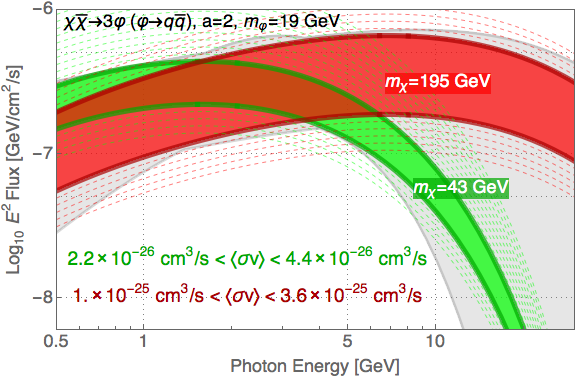}	
\end{center}
\caption{Summary plots for MFV-violating couplings following the figures in the main text. \label{fig:non:MFV}}
\end{figure*}

In the main text we made the point that minimal flavor violation (MFV) suggests that spin-1 mediators couple flavor-universally while spin-0 mediators couple according to the Yukawas. For completeness, we present summary plots of the opposite scenarios in Fig.~\ref{fig:non:MFV}. 

\bibliography{HooperonSpectrumTrimmed}

\end{document}